\documentclass[aps,prl,showpacs]{revtex4-1}

\usepackage{graphicx}
\usepackage{amssymb,amsmath}
\usepackage{natbib}
\usepackage{subfigure}
\usepackage{epstopdf}

\newcommand{\tpr}{\text{Pr}}
\newcommand{\tra}{\text{Ra}}
\newcommand{\tnu}{\text{Nu}}

\newcommand{\tri}{\text{Ri}}

\begin{document}



\title{Logarithmic mean temperature profiles and their connection to plume emissions in 
turbulent Rayleigh-B\'enard convection}
\author{Erwin P. van der Poel$^1$, Rodolfo Ostilla-M\'onico$^1$, Roberto Verzicco$^{2,1}$, Siegfried Grossmann$^{3}$ and Detlef Lohse$^{1}$}
\affiliation{
$^1$Department of Physics, Mesa+ Institute,  and J.\ M.\ Burgers Centre for Fluid Dynamics, University of Twente, 7500 AE Enschede, The Netherlands \\
$^2$Dipartimento di Ingegneria Industriale, University of Rome ``Tor Vergata'', Via del Politecnico 1, Roma 00133, Italy \\
$^3$Fachbereich Physik, Philipps-Universität Marburg, Renthof 6, D-35032 Marburg, Germany}

\date{\today}

\begin{abstract}
Two-dimensional simulations of Rayleigh-B\'enard convection at $\tra = 5\times10^{10}$ show that vertical logarithmic mean temperature profiles can be observed in regions of the boundary layer where thermal plumes are emitted. The profile is logarithmic only in these regions and not in the rest of the boundary layer where it is sheared by the large scale wind and impacted by plumes. In addition, the logarithmic behavior is not visible in the horizontal average. The findings reveal that the temperature profiles are strongly connected to thermal plume emission and support a perception that parts of the boundary layer can be turbulent, while others are not. The transition to the ultimate regime, in which the boundary layers are considered to be fully turbulent,
 can therefore be understood as a gradual increases in fraction of the plume-emitting ('turbulent')
  regions of the boundary layer.
\end{abstract}

\maketitle


Rayleigh-B\'enard (RB) convection consists of a fluid heated from below and cooled from above \cite{sig94,kad01,ahl09,loh10,chi12,xia13}. It is commonly used to model natural convection because of its simplicity and ability to reproduce most of the interesting phenomena. The applications of RB convection range from astrophysics and geophysics to industry. Using the Boussinesq approximation, the control parameters of the system are the non-dimensional temperature difference, i.e. the Rayleigh number $\tra = g\beta\Delta L^3 / (\nu\kappa) $, the Prandtl number of the fluid $\tpr = \nu/\kappa$, and the aspect-ratio $\Gamma = D/L$, where $L$ is the height of the sample and $D$ its width, $g$ the gravitational acceleration, $\beta$ is the thermal expansion coefficient and $\nu$ and $\kappa$ the kinematic viscosity and the thermal diffusivity, respectively. $u$ is the velocity normalized by the freefall velocity $\sqrt{g\beta\Delta L}$, $t$ is the time normalized by the freefall time $\sqrt{L / (g\beta\Delta)}$ and $\theta$ is the temperature normalized by $\Delta$, the temperature difference between top and bottom plate, and shifted such that $\theta$ is in the range $0 \leq \theta \leq 1$.

Most theories assume that from a critical Rayleigh number $\tra_c$ onwards, the boundary layer (BL) transitions from laminar to turbulent \cite{mal54,kra62,gro00,gro01,gro11} and base their scaling laws on either a time- and space averaged laminar \cite{mal54,gro00} or turbulent \cite{kra62,spi71,shr90,sig94,gro00,gro01,gro11} BL profile. In the regime of laminar BLs, the mean temperature profile is assumed to be of Prandtl-Blasius-Pohlhausen (PBP) type \cite{pra04,ll87,sch00,gro04} while in the turbulent case a logarithmic (log) profile is expected \cite{kra62,gro12}, analog to turbulent temperature boundary layers in wall turbulence \cite{per82}. Recently, a BL equation that includes both laminar and turbulent contributions has been developed \cite{sch15}. For $\tra < \tra_c$, BL profile measurements reveal a profile that is similar to the expected PBP profile when measured at horizontal regions where the flow conditions most closely match the assumptions required to analytically obtain the PBP profile \cite{zho10,zho10b,ste12,poe13}. However, the lateral dependence is strong \cite{zho10b} and deviations from the PBP profile are observed \cite{wag12,shi12,sch12}.

\begin{figure*}
\centering
\includegraphics[trim=0 0 0 0,clip=true,width=0.85\textwidth]{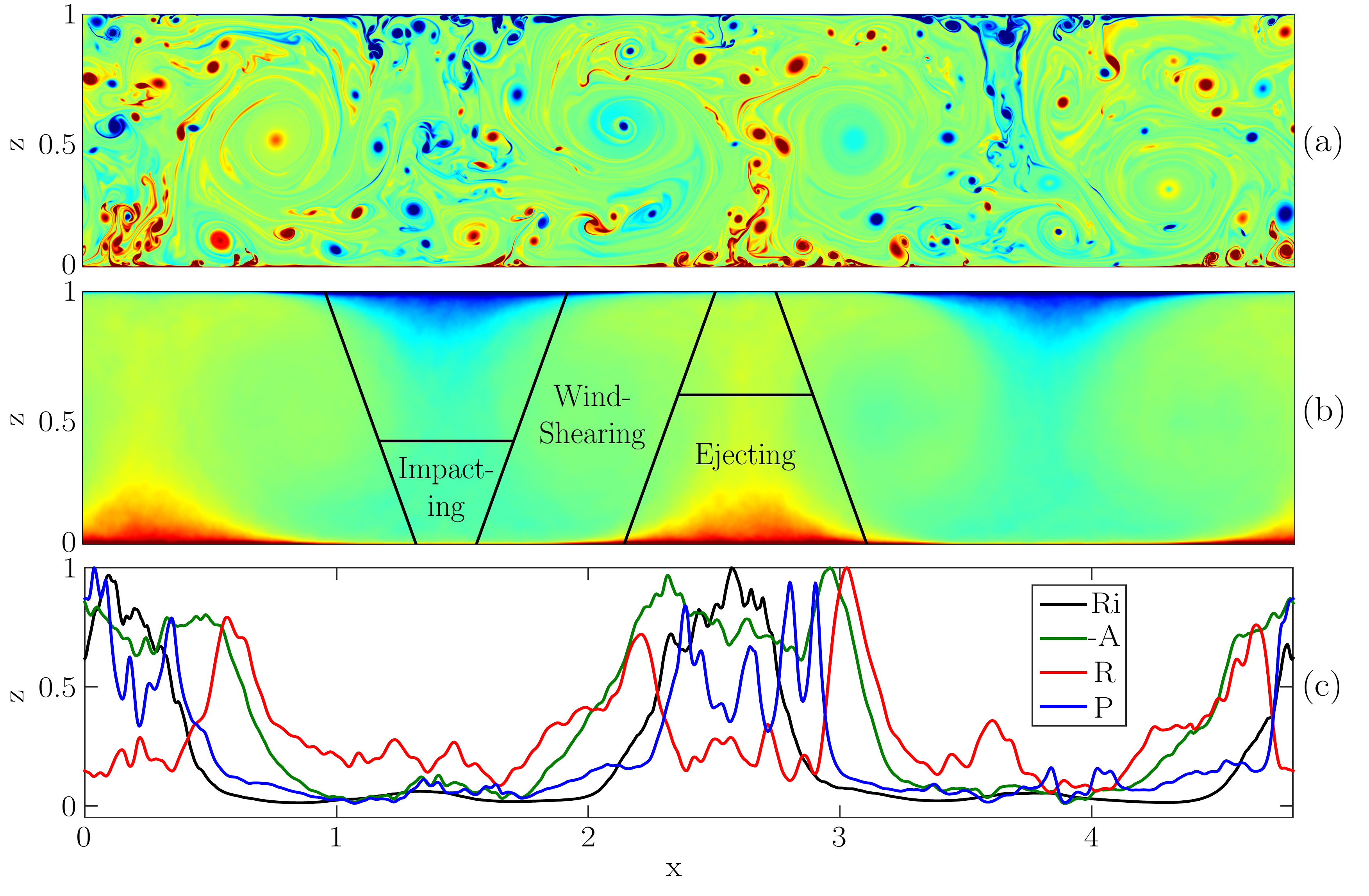}
\caption{a) Instanteneous and b) time-averaged temperature fields. Examples of the three different regions are indicated in the bottom panel. In the panels red and blue indicate hot and cold fluid, respectively. c) The Richardson number Ri, the log amplitude $-A$, the fit residual $R$ and the log quality $P$ as a function of the horizontal coordinate $x$. The plotted quantities are normalized by their maxima. It can be seen that there is a strong correlation between plume ejecting regions and logarithmic mean temperature profiles.}
\label{fig:snaps}
\end{figure*} 

Lateral local mean temperature log profiles have been found in experiments and direct numerical simulations (DNS) down to a Rayleigh number of $\tra = 10^{12}$ \cite{ahl12c,ahl14,wei14}; a Ra where the BL is expected to more closely resemble a laminar profile instead of a logarithmic (turbulent) profile. The log profiles were found near the sidewall in the experiments and numerics, with the log amplitude and fit quality decreasing towards the center of the cell. In cylindrical setups with $\Gamma = \mathcal{O}(1)$ the plumes are commonly ejected from the boundary layer close to the sidewall, resulting in a larger heat flux in the corners \cite{shi07}. The finite size of the cylindrical setup forces the thermal plumes to be emitted near the sidewall due to the large scale circulation of system size. The log profiles could either be an effect of the sidewall or the plume emission or both, as momentum is injected into the RB flow either by the no-slip wall or the thermal plumes. In order to separate these effects, we avoid sidewalls by using lateral periodicity and thereby focus only on the contribution to the logarithmic profile by the thermal plumes. Furthermore, the horizontal plates are impermeable, no-slip and isothermal. Two-dimensional simulations are used to aid visualization and identification of plume emission locations. Even though two-dimensional (2D) RB differs from three-dimensional (3D) RB in terms of integral quantities for finite Pr \cite{sch04,poe13}, the theoretical arguments for logarithmic profiles are not specific to 3D RB. Furthermore, the two-dimensional domain is more suitable to study the horizontal dependence of the boundary layer profile than the three-dimensional domain. The locations of plume emission on the one hand and large scale circulations on the other hand are more straightforwardly identified. In addition, converging local 3D statistics of a sufficiently turbulent flow, which cannot be averaged over any spatial dimension, remains unfeasible to this date due to the required computational cost.


The data are obtained from an energy conserving second order finite difference method \cite{ver99,ver03} converted to 2D. The number of grid points in the used $\tra = 5\times10^{10}$, $\tpr = 1$, aspect-ratio $\Gamma = 4.8$ DNS is $12800 \times 2650$ with grid point clustering at the steep gradients near the horizontal plates. The aspect-ratio is $\Gamma = 4.8$ as this stabilizes the large scale rolls in the system. For lower $\Gamma = 2$ and comparable $\tra = 10^{10}$ it is found that the heat flux approximates the heat flux at infinite aspect-ratio \cite{joh09}. However, this aspect-ratio is insufficient for the analysis presented here, as the horizontal organization of the flow is not static at $\Gamma = 2$. In order to average in time, the temperature conditioned in a plume ejecting region, the flow organization is required to be static. The total averaging time is 550 freefall time units. As the local averages take a very long time to converge, a horizontal moving average with a window of $0.016L$ is used.

\begin{figure}
\centering
\includegraphics[width=0.49\textwidth]{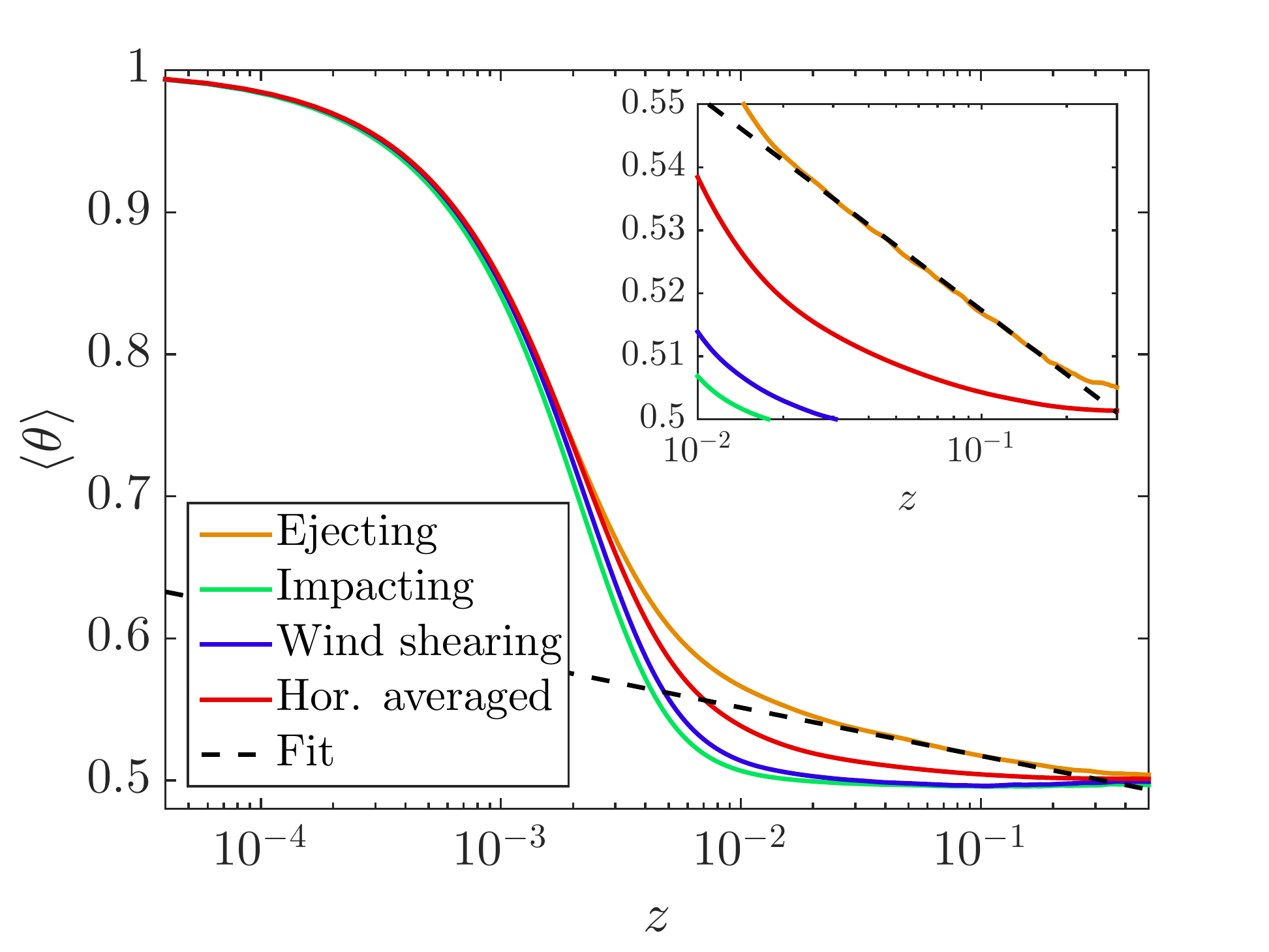}
\caption{Mean temperature profiles for three distinct boundary layer regions and horizontally averaged. The black dashed line is a fit to the logarithmic part of the profile. There is a logarithmic like profile visible in the ejecting mean temperature profile. The region is approximately one decade in width; namely for $0.02 \leq z \leq 0.2$. A zoom can be seen in the inset.}
\label{fig:profiles}
\end{figure} 

It is common in RB theories to assume that there is no horizontal dependence. RB flow can then be conceptually decomposed into boundary layers and a bulk \cite{gro00}. Some authors include an additional mixing zone as an intermediate region \cite{cas89}. However, taking a perspective from channel or pipe flows, where the complete domain can be considered to be a BL, the RB BL corresponds to the inner BL region and the bulk to the outer BL region. This is because in fully developed channel and pipe flow the complete domain is considered to be influenced by the boundary. In addition, these systems are statistically homogeneous in directions parallel to the boundary. In RB flow, this is not the case as it is strongly horizontally (and vertically) inhomogeneous due to the self-organization into large scale rolls. The result of this inhomogeneity is that different regions, which are separated horizontally, have very distinct dynamics. We divide the boundary layer horizontally into three regions, namely {\it ejecting}, {\it impacting} and {\it wind shearing} \cite{ost14}. In the ejecting region thermal plumes are emitted while in the impacting region the wind and plumes from the opposite boundary impact the BL. In between is the wind shearing region that is sheared by the large scale wind. The velocity is predominantly vertical in the ejecting and impacting region while in the wind shearing it is horizontal. These regions are present in both the bottom and top boundary layer, albeit on different horizontal locations. The ejecting regions of the top boundary are opposite of the impacting regions of the bottom boundary and vice versa. In figure \ref{fig:snaps}a the instantaneous temperature field can be seen. Large scale rolls and small scale plumes can be distinguished. As the rolls are nearly stationary, the roll structure and the plume emission spots can clearly be observed in the mean temperature field in figure \ref{fig:snaps}b. The three horizontal regions are sketched on top of this figure, where it can be seen that the ejecting region for one BL is similar to the impacting region for the other BL with respect to the horizontal position. Using the snapshot in \ref{fig:snaps}a, it is apparent that the impacting region is qualitatively different than the wind sheared region and that it also has a smaller width than the ejecting region. 

In order to connect these regions to the logarithmic profiles, the regions must be quantified. For this we use a local Richardson number, which has also been used in forced convection to identify upwelling plumes \cite{tow80}. Here we use a local Richardson number of the form $\tri (x) = \frac{|\langle \theta \rangle_t - \langle \theta \rangle_{x,z,t}|}{ \langle u_x^2 \rangle_t }$. The horizontal $x$ and vertical $z$ coordinates are normalized by $L$. Ri is evaluated at the edge of the thermal boundary layer, which is defined as $\lambda_\theta \equiv L/(2\tnu)$, where $\tnu = \sqrt{\tra\tpr}\langle u_z \theta \rangle_{x,t} - \langle \partial_z \theta \rangle_{x,t}$. In our case $\langle \theta \rangle_{x,z,t} = 1/2$. The Richardson number relates the potential energy with the kinetic energy, commonly used to express the importance of natural convection in relation to forced convection. Using the observation that plumes are predominantly emitted from the BL in regions of low shear, we can use a high Ri to indicate plume upwelling regions and a low Ri for the regions that are sheared by the large scale wind. It must be noted that this does not provide a clear distinction between impacting and wind-shearing regions. However, this is unnecessary here, as we only compare ejecting regions with the other two.

Figure \ref{fig:profiles} shows the respective boundary layer profile for each of the three regions, revealing a logarithmic like profile for the ejecting region. The profiles for the impacting and wind sheared regions are very similar to each other and do not show a clear logarithmic behavior. In addition, the horizontally averaged profile does not show a logarithm either. This is expected considering the studied Ra, as this Ra is not in the so-called ultimate regime that has fully turbulent, logarithmic, boundary layers \cite{gro00,gro01,gro11,ste13}. In previous experiments \cite{ahl12c} the profiles were fitted with the equation $\langle \theta \rangle_t = A~log(z)+B$ for a range of $z \in[0.01,0.1]$. We use this equation but change the range to $z \in[0.02,0.2]$ as it more accurately delimits the logarithmic region for this Ra. We connect the log profile to the thermal plume hotspots through $\tri$ in figure \ref{fig:snaps}c. Here Ri, $-A$, $P$ and the fit residual $R$ are plotted as a function of $x$. The residual $R$ is the Euclidean norm of the difference between the fit and the data. The quantity $P$ is defined as $P = -|A|/R$ and signifies the "log-ness" of the profile. It is a \textit{goodness of fit} parameter. Ri is evaluated at the edge of the thermal BL. All the plotted quantities are normalized by their maxima. The log amplitude $-A$ indicates the amplitude of the fitted log. This value is low in the sheared and impacting region and high in both the upwelling regions. A low residual indicates a good fit. It must be noted that fitting a constant value with a log gives a low residual like in the sheared and impacting regions. However, the log amplitude $-A$ reveals that in this regions there is a negligible log amplitude. The highest $-A$ is connected to a high residual, indicating that the log fit is not good. High amplitudes and low residuals, i.e. good log fits with strong log dependence, are found in the ejecting regions and it becomes clear that there is a strong correlation between the good and strong log profiles and $\tri$, indicating that the log profiles are connected to the plume hotspots and that these logarithmic profiles can be present without sidewalls. The value of $P$ is high throughout the ejecting regions, but has distinct peaks near the edges of these regions. 

\begin{figure}
\centering
\subfigure{\includegraphics[trim=0 55 30 10,clip=true,width=0.49\textwidth]{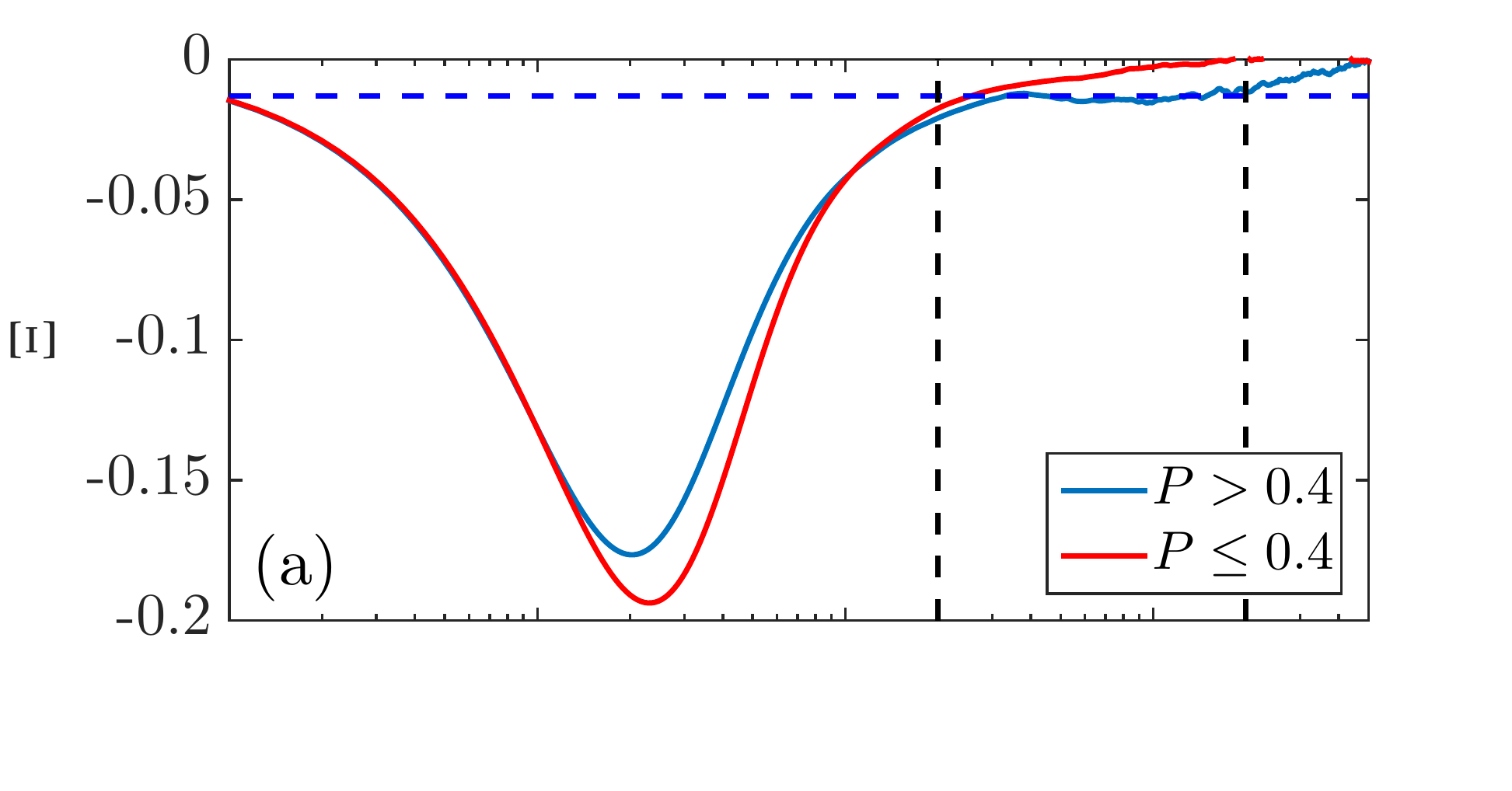}\label{fig:diag}}
\subfigure{\includegraphics[trim=0 5 30 0,clip=true,width=0.49\textwidth]{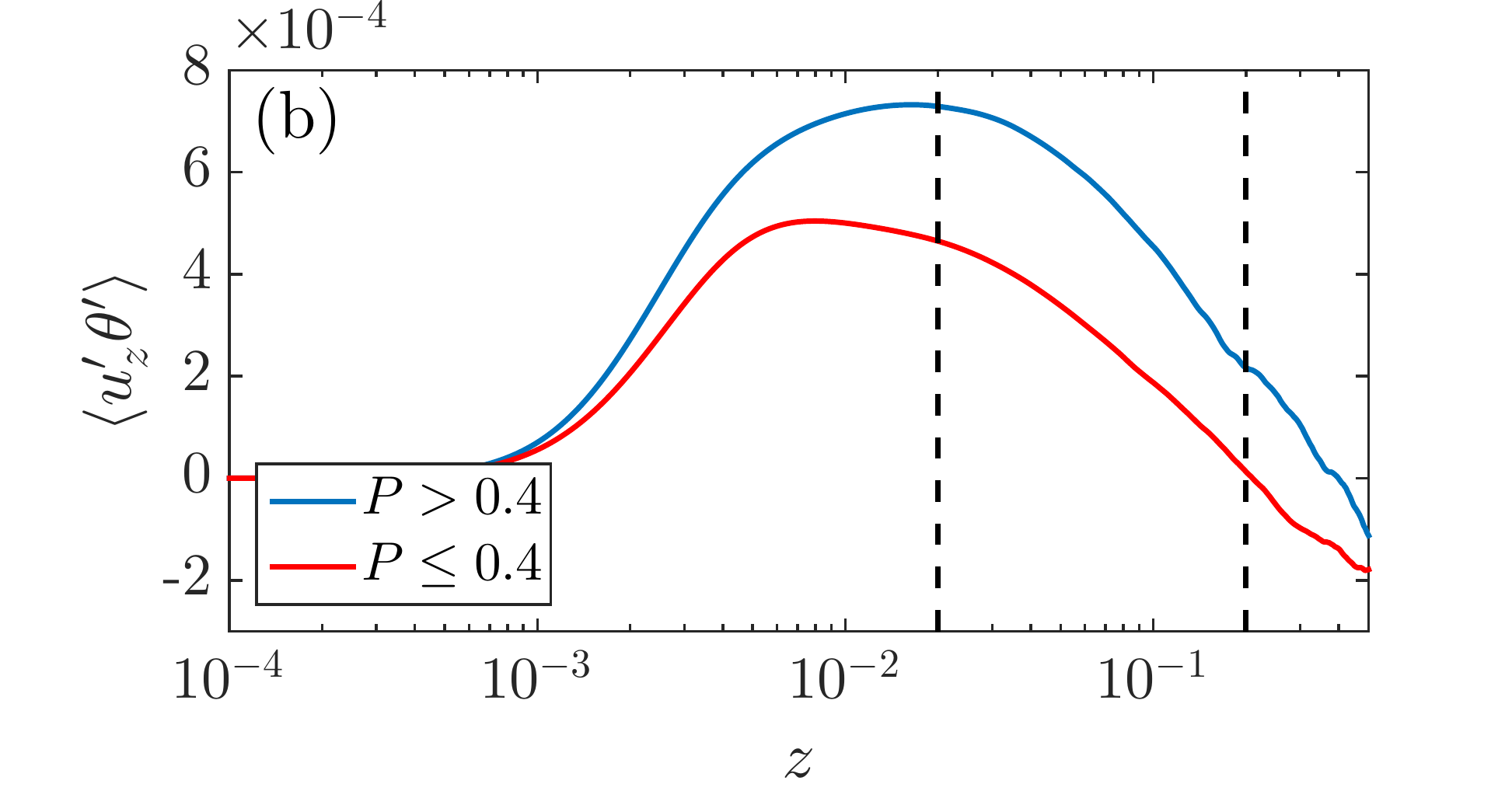}\label{fig:densq3fme}}
\caption{a) Logarithmic diagnostic function $\Xi$ as a function of $z$ for horizontal regions where $P$ is smaller and larger than $0.4$. This function is constant if $\langle \theta \rangle$ is logarithmic in $z$. This is the case in the one-decade log regime between $0.02 \leq z \leq 0.2$, delimited by the dashed black lines. The blue dashed line indicates the logarithmic fit from figure \ref{fig:profiles}. b) $\langle u_z' \theta' \rangle_t$ as a function of $z$ for horizontal regions where $P$ is smaller and larger than $0.4$. The statistics used in these plots are extracted from both the top and bottom boundary layer.}
\end{figure} 

The logarithmic profile can be tested using a logarithmic diagnostic function, which is commonly used in wall bounded turbulence \cite{jim07}. The function in this case for temperature is $\Xi(z) = z\frac{\partial \langle \theta \rangle_t}{\partial z}$. For a logarithmic $\langle \theta \rangle(z)$ profile, this function is constant and can therefore be used to diagnose the logarithmic profile. Because of the required derivative, this function is difficult to converge in both experiments and numerics. In figure \ref{fig:diag} the function is plotted separated for ejecting regions and the rest of the domain. The ejecting region is separated by conditioning on $P$. It appears that using $P > 0.4$ cleanly separates the profile into a logarithmic and a non-logarithmic part. The function $\Xi(z)$ is constant for approximately one decade for $P > 0.4$, while it is clearly not for $P \leq 0.4$. In addition, figure \ref{fig:diag} reveals a typical value for A of $-0.01$.

In turbulent flows the temperature gradient hypothesis can be used to obtain a logarithmic profile for the temperature \cite{gro12}. It states: $\langle u_z' \theta' \rangle_t \hat{=} -\kappa_{turb}(z) \frac{\partial \langle \theta \rangle_t}{\partial z}$, where $\kappa_{turb}(z)$ is the eddy diffusivity, which dominates the heat transfer in case of turbulence. The eddy diffusivity is approximated to depend linearly on the distance from the wall $z$. In that case, the term $\langle u_z' \theta' \rangle_t$ has to be constant throughout the logarithmic temperature layer, since $\frac{\partial \langle \theta \rangle_t}{\partial z} \sim z^{-1}$ as was shown in figure \ref{fig:diag}. The conditioned $\langle u_z' \theta' \rangle_t$ is shown in figure \ref{fig:densq3fme}. This term is time-averaged and averaged over $x$ conditioned on the value of $P$. For the non-logarithmic regions of $P \leq 0.5$, it can be seen that $\langle u_z' \theta' \rangle_t$ is not constant. This is expected for a non-logarithmic mean temperature profile. In case of a logarithmic profile with $\langle u_z' \theta' \rangle_t$ only depending on $z$, it is expected that $\langle u_z' \theta' \rangle_t$ is constant. However, this can hardly be distinguished in figure \ref{fig:densq3fme} for $P > 0.4$, even though the curvature of $\langle u_z' \theta' \rangle_t$ is much less than for $P \leq 0.4$. This observation is very similar to logarithmic profiles in channels \cite{ber14}, which indicates that also in systems where there is horizontal homogeneity this quantity is not exactly constant throughout the logarithmic layer. In addition, in both channel flow and here, the lower curvature of $\langle u_z' \theta' \rangle_t$ is in a region lower in $z$ than the logarithmic layer. This offset seems to be stronger in RB flow, which might be due to the inherent horizontal dependence of the boundary layer profile in this flow. There is horizontal heat flux locally and the common BL approximation $|\partial \theta / \partial x| \ll |\partial \theta / \partial y|$ is not valid in the logarithmic regions. Between $z = 0.01$ and $z = 0.1$ these quantities are even of similar magnitude. 

The flow topology of Rayleigh-B\'enard convection can be divided horizontally into three distinct regions, namely ejecting and impacting regions, where plumes are ejected from the corresponding boundary and impact from the opposite boundary, respectively. In between exists a wind-sheared region, where the large scale wind shears the boundary layer and the flow is predominantly horizontal close to the plates. The profiles of the mean temperature differ stongly between these regions, which explains the radial dependence of the profiles found in cylindrical experiments \cite{ahl12c,ahl14,wei14}. In these experiments, the amplitude and the quality of the logarithmic fit to the data decreases from the sidewall to the central region of the domain. By removing the sidewalls completely, we have shown here that it is the presence of the thermal plumes that result in these profiles. Namely, in a low aspect-ratio cylindrical setup, the thermal plumes move more closely to the sidewall than to the center of the cell. The mean log profiles are connected to the plume hotspots. This signifies that log profiles can and do locally exist in the classical regime and that in this regime the boundary layer profile is not laminar over its full range from certain Ra onwards. We argue that for the studied $\tra = 5\times10^{10}$ the ejecting region has on average a turbulent BL and both the sheared and impacting regions are laminar. The sheared region has a profile that closely resembles a laminar Pohlhausen profile \cite{gro04,zho10,zho10b,ste12,poe13} and the impacting region deviates slightly from this. 

The presence of logarithmic profiles in the classical regime can be reconciled with the ultimate regime by speculating that as a function of Ra the relative size of the hotspot region grows until the full BL is a hotspot and thus has a logarithmic mean profile. Thermal plume related transitions have been seen before \cite{pro91} and it is likely that the transition to the ultimate regime can be triggered by thermal plumes and their local fluctuations instead of a large scale shear. In future work we plan to study the width of the logarithmic region as a function of Rayleigh number to see if there is indeed a continuously growing turbulent boundary layer region up to the possible ultimate regime.

{\it Acknowledgments:} The work was supported by the Foundation for Fundamental Research on Matter (FOM) and the Dutch organization for Scientific Research (NWO).

\bibliography{2dlog.bbl}

\begin{thebibliography}{44}%
\makeatletter
\providecommand \@ifxundefined [1]{%
 \@ifx{#1\undefined}
}%
\providecommand \@ifnum [1]{%
 \ifnum #1\expandafter \@firstoftwo
 \else \expandafter \@secondoftwo
 \fi
}%
\providecommand \@ifx [1]{%
 \ifx #1\expandafter \@firstoftwo
 \else \expandafter \@secondoftwo
 \fi
}%
\providecommand \natexlab [1]{#1}%
\providecommand \enquote  [1]{``#1''}%
\providecommand \bibnamefont  [1]{#1}%
\providecommand \bibfnamefont [1]{#1}%
\providecommand \citenamefont [1]{#1}%
\providecommand \href@noop [0]{\@secondoftwo}%
\providecommand \href [0]{\begingroup \@sanitize@url \@href}%
\providecommand \@href[1]{\@@startlink{#1}\@@href}%
\providecommand \@@href[1]{\endgroup#1\@@endlink}%
\providecommand \@sanitize@url [0]{\catcode `\\12\catcode `\$12\catcode
  `\&12\catcode `\#12\catcode `\^12\catcode `\_12\catcode `\%12\relax}%
\providecommand \@@startlink[1]{}%
\providecommand \@@endlink[0]{}%
\providecommand \url  [0]{\begingroup\@sanitize@url \@url }%
\providecommand \@url [1]{\endgroup\@href {#1}{\urlprefix }}%
\providecommand \urlprefix  [0]{URL }%
\providecommand \Eprint [0]{\href }%
\providecommand \doibase [0]{http://dx.doi.org/}%
\providecommand \selectlanguage [0]{\@gobble}%
\providecommand \bibinfo  [0]{\@secondoftwo}%
\providecommand \bibfield  [0]{\@secondoftwo}%
\providecommand \translation [1]{[#1]}%
\providecommand \BibitemOpen [0]{}%
\providecommand \bibitemStop [0]{}%
\providecommand \bibitemNoStop [0]{.\EOS\space}%
\providecommand \EOS [0]{\spacefactor3000\relax}%
\providecommand \BibitemShut  [1]{\csname bibitem#1\endcsname}%
\let\auto@bib@innerbib\@empty
\bibitem [{\citenamefont {Siggia}(1994)}]{sig94}%
  \BibitemOpen
  \bibfield  {author} {\bibinfo {author} {\bibfnamefont {E.~D.}\ \bibnamefont
  {Siggia}},\ }\href@noop {} {\bibfield  {journal} {\bibinfo  {journal} {Annu.
  Rev. Fluid Mech.}\ }\textbf {\bibinfo {volume} {26}},\ \bibinfo {pages} {137}
  (\bibinfo {year} {1994})}\BibitemShut {NoStop}%
\bibitem [{\citenamefont {Kadanoff}(2001)}]{kad01}%
  \BibitemOpen
  \bibfield  {author} {\bibinfo {author} {\bibfnamefont {L.~P.}\ \bibnamefont
  {Kadanoff}},\ }\href@noop {} {\bibfield  {journal} {\bibinfo  {journal}
  {Phys. Today}\ }\textbf {\bibinfo {volume} {54}},\ \bibinfo {pages} {34}
  (\bibinfo {year} {2001})}\BibitemShut {NoStop}%
\bibitem [{\citenamefont {Ahlers}\ \emph {et~al.}(2009)\citenamefont {Ahlers},
  \citenamefont {Grossmann},\ and\ \citenamefont {Lohse}}]{ahl09}%
  \BibitemOpen
  \bibfield  {author} {\bibinfo {author} {\bibfnamefont {G.}~\bibnamefont
  {Ahlers}}, \bibinfo {author} {\bibfnamefont {S.}~\bibnamefont {Grossmann}}, \
  and\ \bibinfo {author} {\bibfnamefont {D.}~\bibnamefont {Lohse}},\
  }\href@noop {} {\bibfield  {journal} {\bibinfo  {journal} {Rev. Mod. Phys.}\
  }\textbf {\bibinfo {volume} {81}},\ \bibinfo {pages} {503} (\bibinfo {year}
  {2009})}\BibitemShut {NoStop}%
\bibitem [{\citenamefont {Lohse}\ and\ \citenamefont {Xia}(2010)}]{loh10}%
  \BibitemOpen
  \bibfield  {author} {\bibinfo {author} {\bibfnamefont {D.}~\bibnamefont
  {Lohse}}\ and\ \bibinfo {author} {\bibfnamefont {K.~Q.}\ \bibnamefont
  {Xia}},\ }\href@noop {} {\bibfield  {journal} {\bibinfo  {journal} {Annu.
  Rev. Fluid Mech.}\ }\textbf {\bibinfo {volume} {42}},\ \bibinfo {pages} {335}
  (\bibinfo {year} {2010})}\BibitemShut {NoStop}%
\bibitem [{\citenamefont {Chill\`a}\ and\ \citenamefont
  {Schumacher}(2012)}]{chi12}%
  \BibitemOpen
  \bibfield  {author} {\bibinfo {author} {\bibfnamefont {F.}~\bibnamefont
  {Chill\`a}}\ and\ \bibinfo {author} {\bibfnamefont {J.}~\bibnamefont
  {Schumacher}},\ }\href@noop {} {\bibfield  {journal} {\bibinfo  {journal}
  {Eur. Phys. J. E}\ }\textbf {\bibinfo {volume} {35}},\ \bibinfo {pages} {85}
  (\bibinfo {year} {2012})}\BibitemShut {NoStop}%
\bibitem [{\citenamefont {Xia}(2013)}]{xia13}%
  \BibitemOpen
  \bibfield  {author} {\bibinfo {author} {\bibfnamefont {K.-Q.}\ \bibnamefont
  {Xia}},\ }\href@noop {} {\bibfield  {journal} {\bibinfo  {journal} {Appl.
  Mech. Lett.}\ }\textbf {\bibinfo {volume} {3}},\ \bibinfo {pages} {052001}
  (\bibinfo {year} {2013})}\BibitemShut {NoStop}%
\bibitem [{\citenamefont {Malkus}(1954)}]{mal54}%
  \BibitemOpen
  \bibfield  {author} {\bibinfo {author} {\bibfnamefont {M.~V.~R.}\
  \bibnamefont {Malkus}},\ }\href@noop {} {\bibfield  {journal} {\bibinfo
  {journal} {Proc. R. Soc. London A}\ }\textbf {\bibinfo {volume} {225}},\
  \bibinfo {pages} {196} (\bibinfo {year} {1954})}\BibitemShut {NoStop}%
\bibitem [{\citenamefont {Kraichnan}(1962)}]{kra62}%
  \BibitemOpen
  \bibfield  {author} {\bibinfo {author} {\bibfnamefont {R.~H.}\ \bibnamefont
  {Kraichnan}},\ }\href@noop {} {\bibfield  {journal} {\bibinfo  {journal}
  {Phys. Fluids}\ }\textbf {\bibinfo {volume} {5}},\ \bibinfo {pages} {1374}
  (\bibinfo {year} {1962})}\BibitemShut {NoStop}%
\bibitem [{\citenamefont {Grossmann}\ and\ \citenamefont
  {Lohse}(2000)}]{gro00}%
  \BibitemOpen
  \bibfield  {author} {\bibinfo {author} {\bibfnamefont {S.}~\bibnamefont
  {Grossmann}}\ and\ \bibinfo {author} {\bibfnamefont {D.}~\bibnamefont
  {Lohse}},\ }\href@noop {} {\bibfield  {journal} {\bibinfo  {journal} {J.
  Fluid. Mech.}\ }\textbf {\bibinfo {volume} {407}},\ \bibinfo {pages} {27}
  (\bibinfo {year} {2000})}\BibitemShut {NoStop}%
\bibitem [{\citenamefont {Grossmann}\ and\ \citenamefont
  {Lohse}(2001)}]{gro01}%
  \BibitemOpen
  \bibfield  {author} {\bibinfo {author} {\bibfnamefont {S.}~\bibnamefont
  {Grossmann}}\ and\ \bibinfo {author} {\bibfnamefont {D.}~\bibnamefont
  {Lohse}},\ }\href@noop {} {\bibfield  {journal} {\bibinfo  {journal} {Phys.
  Rev. Lett.}\ }\textbf {\bibinfo {volume} {86}},\ \bibinfo {pages} {3316}
  (\bibinfo {year} {2001})}\BibitemShut {NoStop}%
\bibitem [{\citenamefont {Grossmann}\ and\ \citenamefont
  {Lohse}(2011)}]{gro11}%
  \BibitemOpen
  \bibfield  {author} {\bibinfo {author} {\bibfnamefont {S.}~\bibnamefont
  {Grossmann}}\ and\ \bibinfo {author} {\bibfnamefont {D.}~\bibnamefont
  {Lohse}},\ }\href@noop {} {\bibfield  {journal} {\bibinfo  {journal} {Phys.
  Fluids}\ }\textbf {\bibinfo {volume} {23}},\ \bibinfo {pages} {045108}
  (\bibinfo {year} {2011})}\BibitemShut {NoStop}%
\bibitem [{\citenamefont {Spiegel}(1971)}]{spi71}%
  \BibitemOpen
  \bibfield  {author} {\bibinfo {author} {\bibfnamefont {E.~A.}\ \bibnamefont
  {Spiegel}},\ }\href@noop {} {\bibfield  {journal} {\bibinfo  {journal} {Ann.
  Rev. Astron. Astrophys.}\ }\textbf {\bibinfo {volume} {9}},\ \bibinfo {pages}
  {323} (\bibinfo {year} {1971})}\BibitemShut {NoStop}%
\bibitem [{\citenamefont {Shraiman}\ and\ \citenamefont
  {Siggia}(1990)}]{shr90}%
  \BibitemOpen
  \bibfield  {author} {\bibinfo {author} {\bibfnamefont {B.~I.}\ \bibnamefont
  {Shraiman}}\ and\ \bibinfo {author} {\bibfnamefont {E.~D.}\ \bibnamefont
  {Siggia}},\ }\href@noop {} {\bibfield  {journal} {\bibinfo  {journal} {Phys.
  Rev. A}\ }\textbf {\bibinfo {volume} {42}},\ \bibinfo {pages} {3650}
  (\bibinfo {year} {1990})}\BibitemShut {NoStop}%
\bibitem [{\citenamefont {Prandtl}(1905)}]{pra04}%
  \BibitemOpen
  \bibfield  {author} {\bibinfo {author} {\bibfnamefont {L.}~\bibnamefont
  {Prandtl}},\ }in\ \href@noop {} {\emph {\bibinfo {booktitle} {Verhandlungen
  des III. Int. Math. Kongr., Heidelberg, 1904}}}\ (\bibinfo  {publisher}
  {Teubner},\ \bibinfo {address} {Leipzig},\ \bibinfo {year} {1905})\ pp.\
  \bibinfo {pages} {484--491}\BibitemShut {NoStop}%
\bibitem [{\citenamefont {Landau}\ and\ \citenamefont {Lifshitz}(1987)}]{ll87}%
  \BibitemOpen
  \bibfield  {author} {\bibinfo {author} {\bibfnamefont {L.~D.}\ \bibnamefont
  {Landau}}\ and\ \bibinfo {author} {\bibfnamefont {E.~M.}\ \bibnamefont
  {Lifshitz}},\ }\href@noop {} {\emph {\bibinfo {title} {Fluid Mechanics}}}\
  (\bibinfo  {publisher} {Pergamon Press},\ \bibinfo {address} {Oxford},\
  \bibinfo {year} {1987})\BibitemShut {NoStop}%
\bibitem [{\citenamefont {Schlichting}\ and\ \citenamefont
  {Gersten}(2000)}]{sch00}%
  \BibitemOpen
  \bibfield  {author} {\bibinfo {author} {\bibfnamefont {H.}~\bibnamefont
  {Schlichting}}\ and\ \bibinfo {author} {\bibfnamefont {K.}~\bibnamefont
  {Gersten}},\ }\href@noop {} {\emph {\bibinfo {title} {Boundary layer
  theory}}},\ \bibinfo {edition} {8th}\ ed.\ (\bibinfo  {publisher} {Springer
  Verlag},\ \bibinfo {address} {Berlin},\ \bibinfo {year} {2000})\BibitemShut
  {NoStop}%
\bibitem [{\citenamefont {Grossmann}\ and\ \citenamefont
  {Lohse}(2004)}]{gro04}%
  \BibitemOpen
  \bibfield  {author} {\bibinfo {author} {\bibfnamefont {S.}~\bibnamefont
  {Grossmann}}\ and\ \bibinfo {author} {\bibfnamefont {D.}~\bibnamefont
  {Lohse}},\ }\href@noop {} {\bibfield  {journal} {\bibinfo  {journal} {Phys.
  Fluids}\ }\textbf {\bibinfo {volume} {16}},\ \bibinfo {pages} {4462}
  (\bibinfo {year} {2004})}\BibitemShut {NoStop}%
\bibitem [{\citenamefont {Grossmann}\ and\ \citenamefont
  {Lohse}(2012)}]{gro12}%
  \BibitemOpen
  \bibfield  {author} {\bibinfo {author} {\bibfnamefont {S.}~\bibnamefont
  {Grossmann}}\ and\ \bibinfo {author} {\bibfnamefont {D.}~\bibnamefont
  {Lohse}},\ }\href@noop {} {\bibfield  {journal} {\bibinfo  {journal} {Phys.
  Fluids.}\ }\textbf {\bibinfo {volume} {24}},\ \bibinfo {pages} {125103}
  (\bibinfo {year} {2012})}\BibitemShut {NoStop}%
\bibitem [{\citenamefont {Perry}\ and\ \citenamefont {Chong}(1982)}]{per82}%
  \BibitemOpen
  \bibfield  {author} {\bibinfo {author} {\bibfnamefont {A.~E.}\ \bibnamefont
  {Perry}}\ and\ \bibinfo {author} {\bibfnamefont {M.~S.}\ \bibnamefont
  {Chong}},\ }\href@noop {} {\bibfield  {journal} {\bibinfo  {journal} {J.
  Fluid. Mech.}\ }\textbf {\bibinfo {volume} {119}},\ \bibinfo {pages} {173}
  (\bibinfo {year} {1982})}\BibitemShut {NoStop}%
\bibitem [{\citenamefont {Shishkina}\ \emph {et~al.}(2015)\citenamefont
  {Shishkina}, \citenamefont {Horn}, \citenamefont {Wagner},\ and\
  \citenamefont {Ching}}]{sch15}%
  \BibitemOpen
  \bibfield  {author} {\bibinfo {author} {\bibfnamefont {O.}~\bibnamefont
  {Shishkina}}, \bibinfo {author} {\bibfnamefont {S.}~\bibnamefont {Horn}},
  \bibinfo {author} {\bibfnamefont {S.}~\bibnamefont {Wagner}}, \ and\ \bibinfo
  {author} {\bibfnamefont {E.~S.~C.}\ \bibnamefont {Ching}},\ }\href@noop {}
  {\bibfield  {journal} {\bibinfo  {journal} {Phys. Rev. Lett.}\ }\textbf
  {\bibinfo {volume} {114}},\ \bibinfo {pages} {114302} (\bibinfo {year}
  {2015})}\BibitemShut {NoStop}%
\bibitem [{\citenamefont {Zhou}\ and\ \citenamefont {Xia}(2010)}]{zho10}%
  \BibitemOpen
  \bibfield  {author} {\bibinfo {author} {\bibfnamefont {Q.}~\bibnamefont
  {Zhou}}\ and\ \bibinfo {author} {\bibfnamefont {K.-Q.}\ \bibnamefont {Xia}},\
  }\href@noop {} {\bibfield  {journal} {\bibinfo  {journal} {Phys. Rev. Lett.}\
  }\textbf {\bibinfo {volume} {104}},\ \bibinfo {pages} {104301} (\bibinfo
  {year} {2010})}\BibitemShut {NoStop}%
\bibitem [{\citenamefont {Zhou}\ \emph {et~al.}(2010)\citenamefont {Zhou},
  \citenamefont {Stevens}, \citenamefont {Sugiyama}, \citenamefont {Grossmann},
  \citenamefont {Lohse},\ and\ \citenamefont {Xia}}]{zho10b}%
  \BibitemOpen
  \bibfield  {author} {\bibinfo {author} {\bibfnamefont {Q.}~\bibnamefont
  {Zhou}}, \bibinfo {author} {\bibfnamefont {R.~J. A.~M.}\ \bibnamefont
  {Stevens}}, \bibinfo {author} {\bibfnamefont {K.}~\bibnamefont {Sugiyama}},
  \bibinfo {author} {\bibfnamefont {S.}~\bibnamefont {Grossmann}}, \bibinfo
  {author} {\bibfnamefont {D.}~\bibnamefont {Lohse}}, \ and\ \bibinfo {author}
  {\bibfnamefont {K.-Q.}\ \bibnamefont {Xia}},\ }\href@noop {} {\bibfield
  {journal} {\bibinfo  {journal} {J. Fluid. Mech.}\ }\textbf {\bibinfo {volume}
  {664}},\ \bibinfo {pages} {297–312} (\bibinfo {year} {2010})}\BibitemShut
  {NoStop}%
\bibitem [{\citenamefont {Stevens}\ \emph {et~al.}(2012)\citenamefont
  {Stevens}, \citenamefont {Zhou}, \citenamefont {Grossmann}, \citenamefont
  {Verzicco}, \citenamefont {Xia},\ and\ \citenamefont {Lohse}}]{ste12}%
  \BibitemOpen
  \bibfield  {author} {\bibinfo {author} {\bibfnamefont {R.~J. A.~M.}\
  \bibnamefont {Stevens}}, \bibinfo {author} {\bibfnamefont {Q.}~\bibnamefont
  {Zhou}}, \bibinfo {author} {\bibfnamefont {S.}~\bibnamefont {Grossmann}},
  \bibinfo {author} {\bibfnamefont {R.}~\bibnamefont {Verzicco}}, \bibinfo
  {author} {\bibfnamefont {K.-Q.}\ \bibnamefont {Xia}}, \ and\ \bibinfo
  {author} {\bibfnamefont {D.}~\bibnamefont {Lohse}},\ }\href@noop {}
  {\bibfield  {journal} {\bibinfo  {journal} {Phys. Rev. E}\ }\textbf {\bibinfo
  {volume} {85}},\ \bibinfo {pages} {027301} (\bibinfo {year}
  {2012})}\BibitemShut {NoStop}%
\bibitem [{\citenamefont {van~der Poel}\ \emph {et~al.}(2013)\citenamefont
  {van~der Poel}, \citenamefont {Stevens},\ and\ \citenamefont
  {Lohse}}]{poe13}%
  \BibitemOpen
  \bibfield  {author} {\bibinfo {author} {\bibfnamefont {E.~P.}\ \bibnamefont
  {van~der Poel}}, \bibinfo {author} {\bibfnamefont {R.~J. A.~M.}\ \bibnamefont
  {Stevens}}, \ and\ \bibinfo {author} {\bibfnamefont {D.}~\bibnamefont
  {Lohse}},\ }\href@noop {} {\bibfield  {journal} {\bibinfo  {journal} {J.
  Fluid. Mech}\ }\textbf {\bibinfo {volume} {736}},\ \bibinfo {pages} {177}
  (\bibinfo {year} {2013})}\BibitemShut {NoStop}%
\bibitem [{\citenamefont {Wagner}\ \emph {et~al.}(2012)\citenamefont {Wagner},
  \citenamefont {Shishkina},\ and\ \citenamefont {Wagner}}]{wag12}%
  \BibitemOpen
  \bibfield  {author} {\bibinfo {author} {\bibfnamefont {S.}~\bibnamefont
  {Wagner}}, \bibinfo {author} {\bibfnamefont {O.}~\bibnamefont {Shishkina}}, \
  and\ \bibinfo {author} {\bibfnamefont {C.}~\bibnamefont {Wagner}},\
  }\href@noop {} {\bibfield  {journal} {\bibinfo  {journal} {J. Fluid. Mech.}\
  }\textbf {\bibinfo {volume} {697}},\ \bibinfo {pages} {336} (\bibinfo {year}
  {2012})}\BibitemShut {NoStop}%
\bibitem [{\citenamefont {Shi}\ \emph {et~al.}(2012)\citenamefont {Shi},
  \citenamefont {Emran},\ and\ \citenamefont {Schumacher}}]{shi12}%
  \BibitemOpen
  \bibfield  {author} {\bibinfo {author} {\bibfnamefont {N.}~\bibnamefont
  {Shi}}, \bibinfo {author} {\bibfnamefont {M.}~\bibnamefont {Emran}}, \ and\
  \bibinfo {author} {\bibfnamefont {J.}~\bibnamefont {Schumacher}},\
  }\href@noop {} {\bibfield  {journal} {\bibinfo  {journal} {J. Fluid. Mech.}\
  }\textbf {\bibinfo {volume} {706}},\ \bibinfo {pages} {5} (\bibinfo {year}
  {2012})}\BibitemShut {NoStop}%
\bibitem [{\citenamefont {Scheel}\ \emph {et~al.}(2012)\citenamefont {Scheel},
  \citenamefont {Kim},\ and\ \citenamefont {White}}]{sch12}%
  \BibitemOpen
  \bibfield  {author} {\bibinfo {author} {\bibfnamefont {J.}~\bibnamefont
  {Scheel}}, \bibinfo {author} {\bibfnamefont {E.}~\bibnamefont {Kim}}, \ and\
  \bibinfo {author} {\bibfnamefont {K.}~\bibnamefont {White}},\ }\href@noop {}
  {\bibfield  {journal} {\bibinfo  {journal} {J. Fluid. Mech.}\ }\textbf
  {\bibinfo {volume} {711}},\ \bibinfo {pages} {281} (\bibinfo {year}
  {2012})}\BibitemShut {NoStop}%
\bibitem [{\citenamefont {Ahlers}\ \emph {et~al.}(2012)\citenamefont {Ahlers},
  \citenamefont {Bodenschatz}, \citenamefont {Funfschilling}, \citenamefont
  {Grossmann}, \citenamefont {He}, \citenamefont {Lohse}, \citenamefont
  {Stevens},\ and\ \citenamefont {Verzicco}}]{ahl12c}%
  \BibitemOpen
  \bibfield  {author} {\bibinfo {author} {\bibfnamefont {G.}~\bibnamefont
  {Ahlers}}, \bibinfo {author} {\bibfnamefont {E.}~\bibnamefont {Bodenschatz}},
  \bibinfo {author} {\bibfnamefont {D.}~\bibnamefont {Funfschilling}}, \bibinfo
  {author} {\bibfnamefont {S.}~\bibnamefont {Grossmann}}, \bibinfo {author}
  {\bibfnamefont {X.}~\bibnamefont {He}}, \bibinfo {author} {\bibfnamefont
  {D.}~\bibnamefont {Lohse}}, \bibinfo {author} {\bibfnamefont
  {R.}~\bibnamefont {Stevens}}, \ and\ \bibinfo {author} {\bibfnamefont
  {R.}~\bibnamefont {Verzicco}},\ }\href@noop {} {\bibfield  {journal}
  {\bibinfo  {journal} {Phys. Rev. Lett.}\ }\textbf {\bibinfo {volume} {109}},\
  \bibinfo {pages} {114501} (\bibinfo {year} {2012})}\BibitemShut {NoStop}%
\bibitem [{\citenamefont {Ahlers}\ \emph {et~al.}(2014)\citenamefont {Ahlers},
  \citenamefont {Bodenschatz},\ and\ \citenamefont {He}}]{ahl14}%
  \BibitemOpen
  \bibfield  {author} {\bibinfo {author} {\bibfnamefont {G.}~\bibnamefont
  {Ahlers}}, \bibinfo {author} {\bibfnamefont {E.}~\bibnamefont {Bodenschatz}},
  \ and\ \bibinfo {author} {\bibfnamefont {X.}~\bibnamefont {He}},\ }\href@noop
  {} {\bibfield  {journal} {\bibinfo  {journal} {J. Fluid Mech.}\ }\textbf
  {\bibinfo {volume} {758}},\ \bibinfo {pages} {436 } (\bibinfo {year}
  {2014})}\BibitemShut {NoStop}%
\bibitem [{\citenamefont {Wei}\ and\ \citenamefont {Ahlers}(2014)}]{wei14}%
  \BibitemOpen
  \bibfield  {author} {\bibinfo {author} {\bibfnamefont {P.}~\bibnamefont
  {Wei}}\ and\ \bibinfo {author} {\bibfnamefont {G.}~\bibnamefont {Ahlers}},\
  }\href@noop {} {\bibfield  {journal} {\bibinfo  {journal} {J. Fluid Mech.}\
  }\textbf {\bibinfo {volume} {758}},\ \bibinfo {pages} {809 } (\bibinfo {year}
  {2014})}\BibitemShut {NoStop}%
\bibitem [{\citenamefont {Shishkina}\ and\ \citenamefont
  {Wagner}(2007)}]{shi07}%
  \BibitemOpen
  \bibfield  {author} {\bibinfo {author} {\bibfnamefont {O.}~\bibnamefont
  {Shishkina}}\ and\ \bibinfo {author} {\bibfnamefont {C.}~\bibnamefont
  {Wagner}},\ }\href@noop {} {\bibfield  {journal} {\bibinfo  {journal} {Phys
  Fluids.}\ }\textbf {\bibinfo {volume} {19}},\ \bibinfo {pages} {085107}
  (\bibinfo {year} {2007})}\BibitemShut {NoStop}%
\bibitem [{\citenamefont {Schmalzl}\ \emph {et~al.}(2004)\citenamefont
  {Schmalzl}, \citenamefont {Breuer}, \citenamefont {Wessling},\ and\
  \citenamefont {Hansen}}]{sch04}%
  \BibitemOpen
  \bibfield  {author} {\bibinfo {author} {\bibfnamefont {J.}~\bibnamefont
  {Schmalzl}}, \bibinfo {author} {\bibfnamefont {M.}~\bibnamefont {Breuer}},
  \bibinfo {author} {\bibfnamefont {S.}~\bibnamefont {Wessling}}, \ and\
  \bibinfo {author} {\bibfnamefont {U.}~\bibnamefont {Hansen}},\ }\href@noop {}
  {\bibfield  {journal} {\bibinfo  {journal} {Europhys. Lett.}\ }\textbf
  {\bibinfo {volume} {67}},\ \bibinfo {pages} {390} (\bibinfo {year}
  {2004})}\BibitemShut {NoStop}%
\bibitem [{\citenamefont {Verzicco}\ and\ \citenamefont
  {Camussi}(1999)}]{ver99}%
  \BibitemOpen
  \bibfield  {author} {\bibinfo {author} {\bibfnamefont {R.}~\bibnamefont
  {Verzicco}}\ and\ \bibinfo {author} {\bibfnamefont {R.}~\bibnamefont
  {Camussi}},\ }\href@noop {} {\bibfield  {journal} {\bibinfo  {journal} {J.
  Fluid Mech.}\ }\textbf {\bibinfo {volume} {383}},\ \bibinfo {pages} {55}
  (\bibinfo {year} {1999})}\BibitemShut {NoStop}%
\bibitem [{\citenamefont {Verzicco}\ and\ \citenamefont
  {Camussi}(2003)}]{ver03}%
  \BibitemOpen
  \bibfield  {author} {\bibinfo {author} {\bibfnamefont {R.}~\bibnamefont
  {Verzicco}}\ and\ \bibinfo {author} {\bibfnamefont {R.}~\bibnamefont
  {Camussi}},\ }\href@noop {} {\bibfield  {journal} {\bibinfo  {journal} {J.
  Fluid Mech.}\ }\textbf {\bibinfo {volume} {477}},\ \bibinfo {pages} {19}
  (\bibinfo {year} {2003})}\BibitemShut {NoStop}%
\bibitem [{\citenamefont {van~der Poel}\ \emph {et~al.}(2014)\citenamefont
  {van~der Poel}, \citenamefont {Ostilla-M\'{o}nico}, \citenamefont
  {Verzicco},\ and\ \citenamefont {Lohse}}]{poe14}%
  \BibitemOpen
  \bibfield  {author} {\bibinfo {author} {\bibfnamefont {E.~P.}\ \bibnamefont
  {van~der Poel}}, \bibinfo {author} {\bibfnamefont {R.}~\bibnamefont
  {Ostilla-M\'{o}nico}}, \bibinfo {author} {\bibfnamefont {R.}~\bibnamefont
  {Verzicco}}, \ and\ \bibinfo {author} {\bibfnamefont {D.}~\bibnamefont
  {Lohse}},\ }\href@noop {} {\bibfield  {journal} {\bibinfo  {journal} {Phys.
  Rev. E}\ }\textbf {\bibinfo {volume} {90}},\ \bibinfo {pages} {013017}
  (\bibinfo {year} {2014})}\BibitemShut {NoStop}%
\bibitem [{\citenamefont {van~der Poel}\ \emph {et~al.}(2012)\citenamefont
  {van~der Poel}, \citenamefont {Stevens}, \citenamefont {Sugiyama},\ and\
  \citenamefont {Lohse}}]{poe12}%
  \BibitemOpen
  \bibfield  {author} {\bibinfo {author} {\bibfnamefont {E.~P.}\ \bibnamefont
  {van~der Poel}}, \bibinfo {author} {\bibfnamefont {R.~J. A.~M.}\ \bibnamefont
  {Stevens}}, \bibinfo {author} {\bibfnamefont {K.}~\bibnamefont {Sugiyama}}, \
  and\ \bibinfo {author} {\bibfnamefont {D.}~\bibnamefont {Lohse}},\
  }\href@noop {} {\bibfield  {journal} {\bibinfo  {journal} {Phys. Fluids}\
  }\textbf {\bibinfo {volume} {24}},\ \bibinfo {pages} {085104} (\bibinfo
  {year} {2012})}\BibitemShut {NoStop}%
\bibitem [{\citenamefont {Johnston}\ and\ \citenamefont
  {Doering}(2009)}]{joh09}%
  \BibitemOpen
  \bibfield  {author} {\bibinfo {author} {\bibfnamefont {H.}~\bibnamefont
  {Johnston}}\ and\ \bibinfo {author} {\bibfnamefont {C.~R.}\ \bibnamefont
  {Doering}},\ }\href@noop {} {\bibfield  {journal} {\bibinfo  {journal} {Phys.
  Rev. Lett.}\ }\textbf {\bibinfo {volume} {102}},\ \bibinfo {pages} {064501}
  (\bibinfo {year} {2009})}\BibitemShut {NoStop}%
\bibitem [{\citenamefont {Castaing}\ \emph {et~al.}(1989)\citenamefont
  {Castaing}, \citenamefont {Gunaratne}, \citenamefont {Heslot}, \citenamefont
  {Kadanoff}, \citenamefont {Libchaber}, \citenamefont {Thomae}, \citenamefont
  {Wu}, \citenamefont {Zaleski},\ and\ \citenamefont {Zanetti}}]{cas89}%
  \BibitemOpen
  \bibfield  {author} {\bibinfo {author} {\bibfnamefont {B.}~\bibnamefont
  {Castaing}}, \bibinfo {author} {\bibfnamefont {G.}~\bibnamefont {Gunaratne}},
  \bibinfo {author} {\bibfnamefont {F.}~\bibnamefont {Heslot}}, \bibinfo
  {author} {\bibfnamefont {L.}~\bibnamefont {Kadanoff}}, \bibinfo {author}
  {\bibfnamefont {A.}~\bibnamefont {Libchaber}}, \bibinfo {author}
  {\bibfnamefont {S.}~\bibnamefont {Thomae}}, \bibinfo {author} {\bibfnamefont
  {X.~Z.}\ \bibnamefont {Wu}}, \bibinfo {author} {\bibfnamefont
  {S.}~\bibnamefont {Zaleski}}, \ and\ \bibinfo {author} {\bibfnamefont
  {G.}~\bibnamefont {Zanetti}},\ }\href@noop {} {\bibfield  {journal} {\bibinfo
   {journal} {J. Fluid Mech.}\ }\textbf {\bibinfo {volume} {204}},\ \bibinfo
  {pages} {1} (\bibinfo {year} {1989})}\BibitemShut {NoStop}%
\bibitem [{\citenamefont {Ostilla-M\'{o}nico}\ \emph
  {et~al.}(2014)\citenamefont {Ostilla-M\'{o}nico}, \citenamefont {van~der
  Poel}, \citenamefont {Verzicco}, \citenamefont {Grossmann},\ and\
  \citenamefont {Lohse}}]{ost14}%
  \BibitemOpen
  \bibfield  {author} {\bibinfo {author} {\bibfnamefont {R.}~\bibnamefont
  {Ostilla-M\'{o}nico}}, \bibinfo {author} {\bibfnamefont {E.~P.}\ \bibnamefont
  {van~der Poel}}, \bibinfo {author} {\bibfnamefont {R.}~\bibnamefont
  {Verzicco}}, \bibinfo {author} {\bibfnamefont {S.}~\bibnamefont {Grossmann}},
  \ and\ \bibinfo {author} {\bibfnamefont {D.}~\bibnamefont {Lohse}},\
  }\href@noop {} {\bibfield  {journal} {\bibinfo  {journal} {Phys. Fluids}\
  }\textbf {\bibinfo {volume} {26}},\ \bibinfo {pages} {015114} (\bibinfo
  {year} {2014})}\BibitemShut {NoStop}%
\bibitem [{\citenamefont {Townsend}(1980)}]{tow80}%
  \BibitemOpen
  \bibfield  {author} {\bibinfo {author} {\bibfnamefont {A.~A.}\ \bibnamefont
  {Townsend}},\ }\href@noop {} {\emph {\bibinfo {title} {The structure of
  turbulent shear flow}}}\ (\bibinfo  {publisher} {Cambridge University
  Press},\ \bibinfo {year} {1980})\BibitemShut {NoStop}%
\bibitem [{\citenamefont {Stevens}\ \emph {et~al.}(2013)\citenamefont
  {Stevens}, \citenamefont {van~der Poel}, \citenamefont {Grossmann},\ and\
  \citenamefont {Lohse}}]{ste13}%
  \BibitemOpen
  \bibfield  {author} {\bibinfo {author} {\bibfnamefont {R.~J. A.~M.}\
  \bibnamefont {Stevens}}, \bibinfo {author} {\bibfnamefont {E.~P.}\
  \bibnamefont {van~der Poel}}, \bibinfo {author} {\bibfnamefont
  {S.}~\bibnamefont {Grossmann}}, \ and\ \bibinfo {author} {\bibfnamefont
  {D.}~\bibnamefont {Lohse}},\ }\href@noop {} {\bibfield  {journal} {\bibinfo
  {journal} {J. Fluid. Mech.}\ }\textbf {\bibinfo {volume} {730}},\ \bibinfo
  {pages} {295} (\bibinfo {year} {2013})}\BibitemShut {NoStop}%
\bibitem [{\citenamefont {Jim\'{e}nez}\ and\ \citenamefont
  {Moser}(2007)}]{jim07}%
  \BibitemOpen
  \bibfield  {author} {\bibinfo {author} {\bibfnamefont {J.}~\bibnamefont
  {Jim\'{e}nez}}\ and\ \bibinfo {author} {\bibfnamefont {R.~D.}\ \bibnamefont
  {Moser}},\ }\href@noop {} {\bibfield  {journal} {\bibinfo  {journal} {Phil.
  Trans. R. Soc. Lond. A}\ }\textbf {\bibinfo {volume} {365}},\ \bibinfo
  {pages} {715 } (\bibinfo {year} {2007})}\BibitemShut {NoStop}%
\bibitem [{\citenamefont {Bernardini}\ \emph {et~al.}(2014)\citenamefont
  {Bernardini}, \citenamefont {Pirozzoli},\ and\ \citenamefont
  {Orlandi}}]{ber14}%
  \BibitemOpen
  \bibfield  {author} {\bibinfo {author} {\bibfnamefont {M.}~\bibnamefont
  {Bernardini}}, \bibinfo {author} {\bibfnamefont {S.}~\bibnamefont
  {Pirozzoli}}, \ and\ \bibinfo {author} {\bibfnamefont {P.}~\bibnamefont
  {Orlandi}},\ }\href@noop {} {\bibfield  {journal} {\bibinfo  {journal} {J.
  Fluid Mech.}\ }\textbf {\bibinfo {volume} {742}},\ \bibinfo {pages} {171}
  (\bibinfo {year} {2014})}\BibitemShut {NoStop}%
\bibitem [{\citenamefont {Procaccia}\ \emph {et~al.}(1991)\citenamefont
  {Procaccia}, \citenamefont {Ching}, \citenamefont {Constantin}, \citenamefont
  {Kadanoff}, \citenamefont {Libchaber},\ and\ \citenamefont {Wu}}]{pro91}%
  \BibitemOpen
  \bibfield  {author} {\bibinfo {author} {\bibfnamefont {I.}~\bibnamefont
  {Procaccia}}, \bibinfo {author} {\bibfnamefont {E.~S.~C.}\ \bibnamefont
  {Ching}}, \bibinfo {author} {\bibfnamefont {P.}~\bibnamefont {Constantin}},
  \bibinfo {author} {\bibfnamefont {L.~P.}\ \bibnamefont {Kadanoff}}, \bibinfo
  {author} {\bibfnamefont {A.}~\bibnamefont {Libchaber}}, \ and\ \bibinfo
  {author} {\bibfnamefont {X.~Z.}\ \bibnamefont {Wu}},\ }\href@noop {}
  {\bibfield  {journal} {\bibinfo  {journal} {Phys. Rev. A}\ }\textbf {\bibinfo
  {volume} {44}},\ \bibinfo {pages} {8091} (\bibinfo {year}
  {1991})}\BibitemShut {NoStop}%
\end{thebibliography}%

\end{document}